\documentclass[aps,pra,twocolumn,showpacs,superscriptaddress,longbibliography]{revtex4-2}
\usepackage{graphicx} 
\usepackage{standalone}
\usepackage{physics}
\usepackage{amsmath}
\usepackage{graphicx,epstopdf}

\usepackage[flushleft]{threeparttable}
\usepackage{gensymb}
\usepackage{braket}
\newcommand{\be}{\begin{equation}}
\newcommand{\ee}{\end{equation}}
\newcommand{\bea}{\begin{eqnarray}}
\newcommand{\eea}{\end{eqnarray}}
\newcommand{\bse}{\begin{subequations}}
	\newcommand{\ese}{\end{subequations}}

\usepackage{color}
\usepackage[colorlinks,bookmarks=false,citecolor=darkblue,linkcolor=red,urlcolor=blue]{hyperref}

\definecolor{darkred}{rgb}{0.7,0.0,0.0}

\definecolor{darkblue}{rgb}{0,0.02,0.45}

\definecolor{darkgreen}{rgb}{0.02,0.45,0.0}

\definecolor{violet}{rgb}{0.8,0.2,0.6}

\begin{document}

\title{Crystal structure and magnetic properties of spin-$1/2$ frustrated two-leg ladder compounds (C$_4$H$_{14}$N$_2$)Cu$_2X_6$ ($X$= Cl and Br)}

\author{P. Biswal}
\thanks{These authors contributed equally to this work.}
\affiliation{Institute of Physics, Bhubaneswar 751005, India}
\affiliation{Homi Bhabha National Institute, Anushaktinagar, Mumbai 400094, India}
\author{S. Guchhait}
\thanks{These authors contributed equally to this work.}
\affiliation{School of Physics, Indian Institute of Science Education and Research Thiruvananthapuram-695551, Kerala, India}
\author{S. Ghosh}
\thanks{These authors contributed equally to this work.}
\affiliation{Department of Condensed Matter and Materials Physics, S. N. Bose National Centre for Basic Sciences, Kolkata 700106, India}
\author{S. N. Sarangi}
\author{D. Samal}
\affiliation{Institute of Physics, Bhubaneswar 751005, India}
\affiliation{Homi Bhabha National Institute, Anushaktinagar, Mumbai 400094, India}
\author{Diptikant Swain}
\email{diptisscu@gmail.com}
\affiliation{Institute of Chemical Technology–IndianOil Odisha Campus, Bhubaneswar 751013, India}
\author{Manoranjan Kumar}
\affiliation{Department of Condensed Matter and Materials Physics, S. N. Bose National Centre for Basic Sciences, Kolkata 700106, India}
\author{R. Nath}
\email{rnath@iisertvm.ac.in}
\affiliation{School of Physics, Indian Institute of Science Education and Research Thiruvananthapuram-695551, Kerala, India}
\date{\today}

\begin{abstract}
We have successfully synthesized single crystals, solved the crystal structure, and studied the magnetic properties of a new family of copper halides (C$_4$H$_{14}$N$_2$)Cu$_2X_6$ ($X$= Cl, Br). These compounds crystallize in an orthorhombic crystal structure with space group $Pnma$. The crystal structure features Cu$^{2+}$ dimers arranged parallel to each other that makes a zig-zag two-leg ladder-like structure. Further, there exists a diagonal interaction between two adjacent dimers which generates inter-dimer frustration. Both the compounds manifest a singlet ground state with a large gap in the excitation spectrum. Magnetic susceptibility is analyzed in terms of both interacting spin-$1/2$ dimer and two-leg ladder models followed by exact diagonalization calculations. Our theoretical calculations in conjunction with the experimental magnetic susceptibility establish that the spin-lattice can be described well by a frustrated two-leg ladder model with strong rung coupling ($J_0/k_{\rm B} \simeq 116$~K and 300~K), weak leg coupling ($J^{\prime\prime}/k_{\rm B} \simeq 18.6$~K and 105~K), and equally weak diagonal coupling ($J^{\prime }/k_{\rm B} \simeq 23.2$~K and 90~K) for Cl and Br compounds, respectively.
These exchange couplings set the critical fields very high, making them experimentally inaccessible. The correlation function decays exponentially as expected for a gapped spin system. The structural aspects of both the compounds are correlated with their magnetic properties. The calculation of entanglement witness divulges strong entanglement in both the compounds which persists upto high temperatures, even beyond 370~K for the Br compound.
\end{abstract}

\maketitle
\section{Introduction}
In recent days, the low-dimensional spin systems with singlet ground state are pursued rigorously as they manifest interesting field induced quantum phases at low temperatures~\cite{Giamarchi198,Zapf563}. Moreover, the singlet state is considered to be a highly entangled state which has direct bearing on quantum computation and quantum communication~\cite{Connor052302,Vedral1004,Wootters2245}. The singlet ground state can be realized in the spin dimers~\cite{Kofu037206}, alternating spin-chains~\cite{Mukharjee224403,Mukharjee144433}, Haldane chains (integer spin-chains)~\cite{Uchiyama632,ShimizuR9835}, spin-Peierl systems~\cite{Hase3651}, even-leg ladders~\cite{Landee100402,Azuma3463}, frustrated magnets~\cite{Budnik187205} etc.

External magnetic field often acts as a perturbation, which continuously reduces the energy gap between the singlet ($\ket {S,S_z}=\ket {0, 0}$) ground state and the triplet ($\ket {S,S_z}=\ket{1, 1}$) excited states. Above a critical field ($H_{\rm c1}$) when the gap is closed, several intriguing field induced quantum phenomena emerge. To name a few, Bose-Einstein condensation (BEC) of triplons in coupled dimer systems~\cite{Zapf563,Ruegg62,Aczel207203,Ruegg017202,Freitas184426,Hester027201}, Tomonaga-Luttinger Liquid (TLL) in one-dimensional spin chains and spin ladders~\cite{Ozerov241113,Jeong106402,Jeong106404}, magnetization plateaus in interacting dimers~\cite{Kageyama3168,Murugan024451,Uchida054429}, Wigner crystallization~\cite{Horsch076403}, etc have been realized.
The most intricate one being the Shastry-Sutherland lattice that consists of orthogonal dimers embedded in a square lattice. It has an exact dimer product ground state when the ratio between the inter- to intra-dimer couplings is sufficiently low~\cite{Shastry1069}. Upon increasing this ratio, the system goes through a quantum phase transition to a plaquette singlet state followed by an antiferromagnetic phase~\cite{Koga4461}. Applying external field and pressure one can tune the coupling ratio and hence observe a series of quantum phase transitions~\cite{Zayed962,Takigawa067210}.
Indeed, the famous Shastry–Satherland lattice compound SrCu$_2$(BO$_3$)$_2$ featuring orthogonal Cu$^{2+}$ dimers depicts quantized plateaus at $1/8$, $1/4$, and $1/3$ of the magnetization and Wigner crystallization of magnons~\cite{Miyahara3701,Kageyama3168,Onizuka1016,Haravifard11956,Kodama395}. These field induced phases are also observed in several high spin compounds but the quantum effects are more predominant in systems with spin-$1/2$~\cite{Murugan024451,Uchida054429}. Further, the isolated spin dimers with a significant intra-dimer coupling show a large spin-gap, whereas the presence of inter-dimer couplings lead to a drastic reduction in the gap value, making the compounds suitable for high-field studies.

Unlike the transition metal oxides, the metal organic compounds are more suitable for such studies as one can easily tune the inter-dimer and intra-dimer exchange couplings, spin-gap, and the ground state properties by engineering the synthesis conditions and changing the ligand~\cite{Freitas184426}. For instance, the isolated spin-$\frac{1}{2}$ dimers in the metal-organic compound Cu$_{2}$(IPA)$_2$(DMF)(H$_2$O) result a large spin-gap of about $\sim 414$~K~\cite{Thamban255801} while isolated dimers in Cu$_2$Mg$_2$(CO$_3$)(OH)$_6$·2H$_2$O provide a much reduced spin-gap of around $\sim 7$~K~\cite{Lebernegg165127}. The hierarchy of the intra-dimer exchange couplings depends on the interaction path involved, symmetry of the orbitals, bond length, bond angle etc. Further, the inter-dimer coupling which significantly modifies spin-gap can be engineered by an appropriate choice of organic ligands~\cite{Tennant4998,Johnson8123,Arjun174421}. Unfortunately, the database for organic compounds with spin-gap is limited compared to the inorganic counterpart. Interestingly, [Cu$_{2}$(apyhist)$_{2}$Cl$_{2}$](ClO$_{4}$)$_{2}$ (spin-$1/2$) and NiCl$_2$-4SC(NH$_2)_2$
(spin-1) are only two organic compounds reported with small spin-gap and both of them show field induced BEC physics~\cite{Freitas184426,Chiatti094406}.

In this paper, we report in detail the crystal growth, structure and magnetic properties of two iso-structural spin-1/2 strong rung coupled two-leg ladder compounds (C$_4$H$_{14}$N$_2$)Cu$_2X_6$ ($X$ = Cl and Br) with a large spin-gap. Indeed, our experimental magnetic susceptibility data are modeled well by the exact diagonalization (ED) calculations assuming a two-leg ladder model with a diagonal interaction that frustrates the spin-lattice.
It is found that the rung, leg as well as the diagonal couplings for the Br compound are significantly larger as compared to the Cl compound [i.e. $J_0$(Br)/$J_0$(Cl)~$\sim 2.6$, $J^{\prime \prime}$(Br)/$J^{\prime \prime}$(Cl)~$\sim 5.74$, and $J^\prime$(Br)/$J^\prime$(Cl)~$\sim 3.9$]. The relatively large exchange couplings in case of Br compound is attributed to its larger ionic size and more diffused $p$-orbitals which increases the effective coupling between the Cu$^{2+}$ ions. Our work provides a pathway to manipulate the magnetic properties of low-dimensional metal-organic compounds by judiciously changing the halide atom in the magnetically active network.

\section{Techniques}
\begin{figure}
\includegraphics[scale=1]{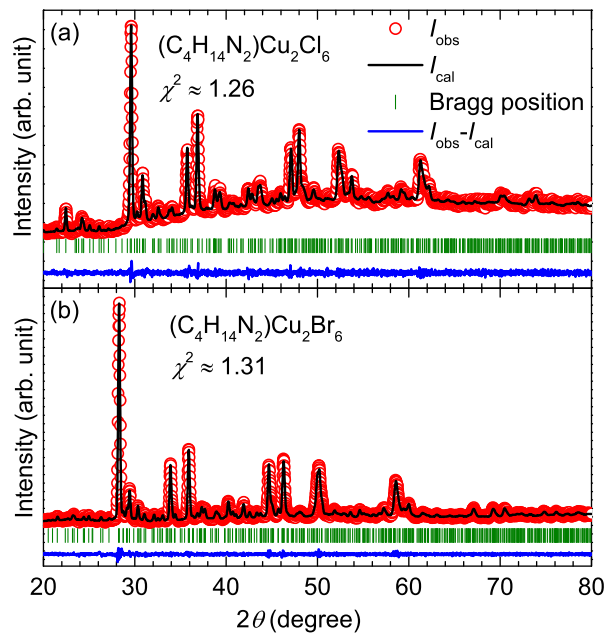}
\caption{Powder XRD patterns (open circles) at room temperature for (a) (C$_4$H$_{14}$N$_2$)Cu$_2$Cl$_6$ and (b) (C$_4$H$_{14}$N$_2$)Cu$_2$Br$_6$. The solid line is the Le Bail fit of the data. The vertical bars indicate the expected Bragg peak positions and the bottom solid line corresponds to the difference between the observed and calculated intensities. 
}
\label{Fig1}
\end{figure}
Single crystals of (C$_4$H$_{14}$N$_2$)Cu$_2$Cl$_6$ and (C$_4$H$_{14}$N$_2$)Cu$_2$Br$_6$ were prepared following the solution evaporation method by using a hot air oven in moderate temperatures. To synthesize (C$_4$H$_{14}$N$_2$)Cu$_2$Cl$_6$, C$_4$H$_{12}$N$_2$ (NN$^{\prime}$-Dimethylethylenediamine) and CuCl$_2$ were added in molar ratio of 1:2 with little excess of HCl. Later, distilled water was added and the solution was heated at 80$^{\degree}$C followed by continuous stirring for complete dissolution of the precursors. The resulting clear solution was kept inside an oven at 45$^{\degree}$C which yielded greenish yellow colored single crystals of (C$_4$H$_{14}$N$_2$)Cu$_2$Cl$_6$ after 5-6 days. Similar procedure was followed to obtain single crystals of (C$_4$H$_{14}$N$_2$)Cu$_2$Br$_6$ by taking the initial constituents C$_4$H$_{12}$N$_2$, CuBr$_2$, and HBr. The only difference is that, the solution was heated at 75$^{\degree}$C instead of 45$^{\degree}$C, which resulted in dark red-colored needle-shaped crystals.

Single crystal x-ray diffraction (XRD) was performed on good quality single crystals at room temperature using a Bruker KAPPA-II machine with a CCD detector and graphite monochromated Mo $K_{\alpha}$ radiation ($\lambda_{\rm avg}=0.71073$~\AA). The data were collected using APEX3 software and reduced with SAINT/XPREP~\cite{Bruker2016apex3}. An empirical absorption correction was done using the SADABS program~\cite{Sheldrick1994}. The structure was solved with direct methods using SHELXT-2018/2~\cite{Sheldrick2015shelxt} and refined by the full matrix least squares on $F^{2}$ using SHELXL-2018/3, respectively~\cite{Sheldrick2018shelxl}. All the hydrogen atoms were placed geometrically and held in the riding atom model for the final refinement. The final refinement included atomic positions for all the atoms, anisotropic thermal parameters for all the nonhydrogen atoms, and isotropic thermal parameters for the hydrogen atoms. The crystal data and detailed information about the structure refinement parameters are listed in Table~\ref{Structural Parameters}. 
To reconfirm the phase purity, a large number of single crystals were crushed into powder and powder XRD measurement was performed at room temperature using a PANalytical (Cu $K_\alpha$ radiation, $\lambda_{\rm avg}=1.5406$~\AA) diffractometer. The powder XRD patterns were analyzed by Le Bail fit using the FULLPROF software package~\cite{Carvajal55}. The initial structural parameters for the fits were taken from the single crystal data (Table~\ref{Structural Parameters}). As shown in Fig.~\ref{Fig1}, all the peaks in the room temperature powder XRD patterns of both compounds could be indexed properly with orthorhombic structure ($Pnma$). The refined lattice parameters and unit cell volume are [$a=15.0583(10)$~\AA, $b=6.0273(5)$~\AA, $c=14.6121(7)$~\AA, and $V_{\rm cell}\simeq1326.22$~\AA$^3$] and [$a=15.7969(11)$~\AA, $b=6.2901(3)$~\AA, $c=14.9536(9)$~\AA, and $V_{\rm cell}\simeq1485.87$~\AA$^3$] for (C$_4$H$_{14}$N$_2$)Cu$_2$Cl$_6$ and (C$_4$H$_{14}$N$_2$)Cu$_2$Br$_6$, respectively. These structural parameters of both the compounds are close to the single crystal data.

Magnetization ($M$) was measured using a SQUID magnetometer (MPMS, Quantum Design) in the temperature range 2~K$\leq T\leq$ 350~K, and in different applied fields ($H$). Isothermal magnetization ($M$ vs $H$) measurement was performed at $T = 2$~K varying the applied field from 0 to 5~T. For this purpose, a bunch of single crystals were aligned and mounted on the sample rod.


The magnetization data are modeled by spin-$1/2$ interacting dimer and two-leg ladder models as well via exact diagonalization (ED) calculations. The ground state of interacting dimers, in general, has a short correlation length. Therefore, small system size ED calculations should be sufficient to obtain a reliable spectrum and thermodynamic properties. We employ the conventional ED method to obtain the full energy spectrum of the system size up to $N=24$ and found that $N=18$ is sufficient to reproduce the experimental magnetization data.

\section{Results}
\subsection{Crystal Structure}
\begin{figure*}
	\includegraphics[width = \linewidth]{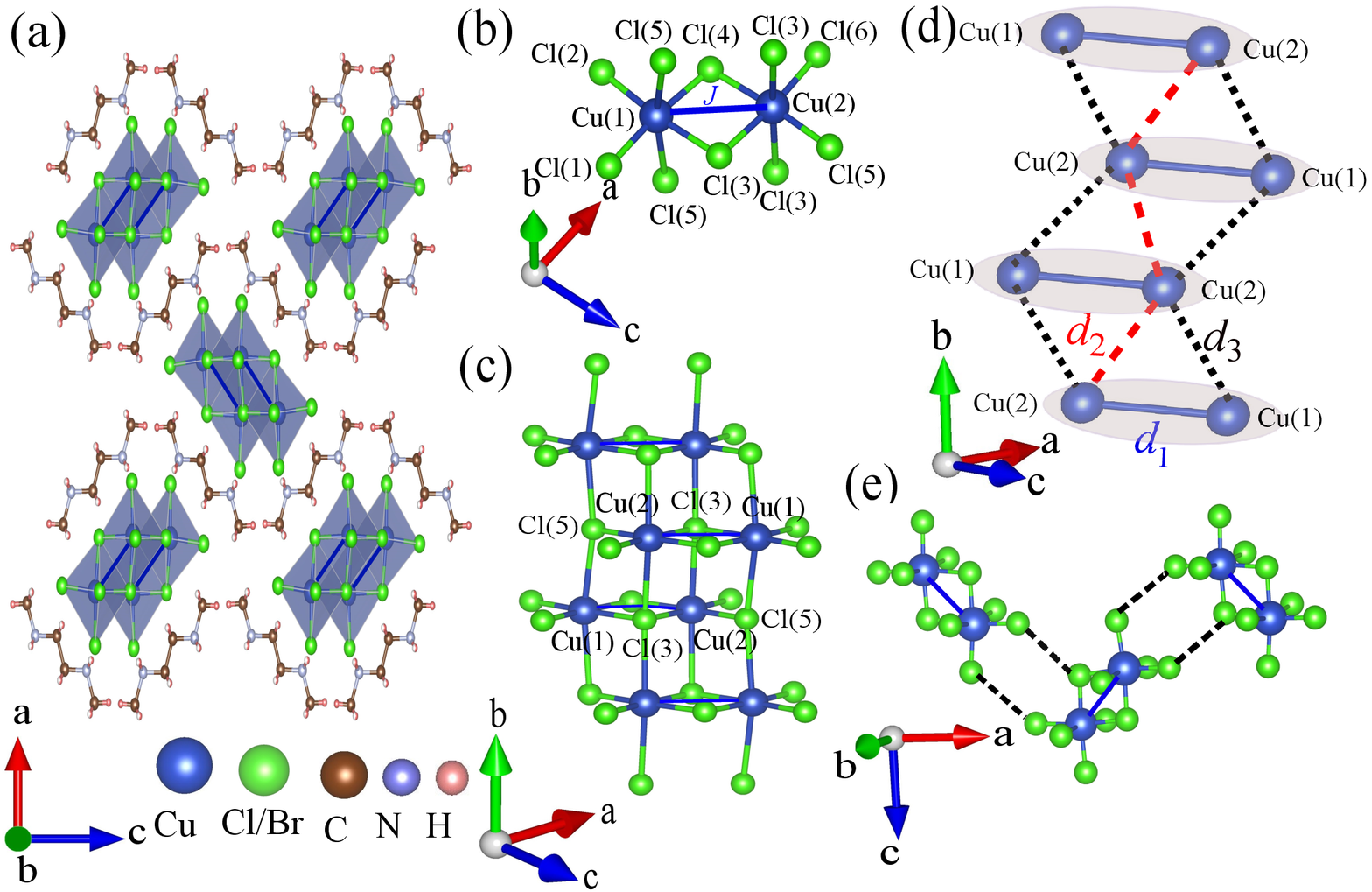}
	\caption{(a) Three dimensional view of the crystal structure of (C$_4$H$_{14}$N$_2$)Cu$_2X_6$ ($X$= Cl, Br) projected in the $ac$-plane. (b) The Cu$_2X_{10}$ ($X$= Cl, Br) dimer unit with intra-dimer exchange coupling $J_0$. (c) Edge sharing of distorted CuX$_6$ octahedra favouring inter-dimer couplings or couplings along the legs and diagonal of the ladder. (d) A sketch of the possible exchange couplings (after removing $X$) highlight the coupled dimers or frustrated two-leg zig-zag ladder structure. (e) Orthogonal dimers from neighboring ladders.}
	\label{Fig2}
\end{figure*}

\begin{table*}
	\setlength{\tabcolsep}{0.5 cm}
	\caption{Structure information of (C$_4$H$_{14}$N$_2$)Cu$_2X_6$ ($X$= Cl, Br) compounds obtained from the single crystal XRD measurements at room temperature.}
	\label{Structural Parameters}
	\begin{tabular}{ccccccc}
		\hline \hline
		\textbf{Crystal data}\\
		Empirical formula & (C$_4$H$_{14}$N$_2$)Cu$_2$Cl$_6$ & (C$_4$H$_{14}$N$_2$)Cu$_2$Br$_6$\\
		Formula weight ($M_r$)&429.95 g/mole& 696.71 g/mole \\	
		Crystal system&orthorhombic&orthorhombic \\
		Space group&$Pnma$&$Pnma$ \\
		$a$ [~\AA] &15.040(1)&15.849(3) \\
		$b$ [~\AA] &6.014(5)&6.3157(5)\\
		$c$[~\AA] &14.593(9)&14.992(3) \\
		$V_{\rm cell}$[~\AA$^3$] &1319.95(17) & 1500.6(5)\\       
		$Z$ & 4 & 4\\ 
		Calculated crystal density ($\rho_{\rm cal}$) & 2.164 mg/mm$^3$ & 3.084 mg/mm$^3$\\
		Absorption coefficient ($\mu$)&4.401 mm$^{-1}$& 18.780 mm$^{-1}$\\ 	
		Crystal size& $0.25\times0.22\times0.17$ mm$^3$& $0.22\times0.18\times0.12$ mm$^3$\\
		\hline	
		\textbf{Data collection}\\
		Temperature~(K)&295(2)&295(2)\\
		Radiation type &Mo$K_{\alpha}$&Mo$K_{\alpha}$\\  
		Wavelength ($\lambda$)&0.71073~\AA&0.71073~\AA\\
		Diffractometer&Bruker KAPPA APEX-II CCD&Bruker KAPPA APEX-II CCD \\	
		$\theta$ range for data collection&2.708$^{\circ}$ to 26.422$^{\circ}$&2.570$^{\circ}$ to 25.462$^{\circ}$\\
		Index ranges & $-18\leq h\leq 18$,& $-19\leq h\leq 19$, &\\
				&$-7\leq k\leq 7$,&$-7\leq k\leq 6$, &\\
				&$-18\leq l\leq 18$&$-18\leq l\leq 18$ &\\
		
			$F$(000) & 884 & 1280\\
				Reflections collected &10316&11768\\
				Independent reflections&1487 [$R_{\rm int} = 0.0462$]&1526 [$R_{\rm int} = 0.0685$] \\
			Data/restraints/parameters&1487/0/85&1526/5/73\\
			Final $R$ indexes, $I\geq 2\sigma(I)$&$R_{1}=0.0261$, $\omega R_{2} =  0.0549$&$R_{1}=0.1005$, $\omega R_{2} =  0.2441$& \\	
			Final $R$ indexes, all data&$R_{1}=0.0389$, $\omega R_{2}=0.0621$&$R_{1}=0.1346$, $\omega R_{2}=0.2904$&  \\
		\hline
		\textbf{Refinement}\\
		Refinement method&Full-matrix least-squares on $F^2$&Full-matrix least-squares on $F^2$\\
		Goodness-of-fit on $F^{2}$&1.123&1.017\\
		\hline\hline 
	\end{tabular}
\end{table*} 
Both compounds are found to have same crystal structure [orthorhombic, $Pnma$ ($D_{2h}^{16}$, No.~62)] with $Z=4$. The detail crystallographic parameters obtained from the single crystal x-ray diffraction analysis such as lattice parameters ($a, b$, and $c$), unit cell volume ($V_{\rm cell}$) etc are summarized in Table~\ref{Structural Parameters}. The atomic positions, bond lengths, bond-angles, and anisotropic atomic displacement parameters are tabulated in Supplementary Material (SM)~\cite{Supplementary}.
The crystal structure obtained from the single crystal XRD is presented in Fig.~\ref{Fig2}. The asymmetric unit of the crystal contains one N,N$^{\prime}$-Dimethylethylenediamonium (NN$^{\prime}$D) cation and one Cu$_2X_6$ ($X$ = Cl, Br) anion where both reside on mirror plane symmetry with an occupancy of 0.50. There are two inequivalent Cu sites and six in-equivalent halide (Cl/Br) sites present in the crystal unit cell. Each Cu atom is coordinated with six halide (Cl/Br) atoms forming a distorted Cu$X_6$ octahedra. The octahedra are highly distorted with significant elongation of Cu-$X$ bond along the apical direction. The calculated distortion parameters are given in SM~\cite{Supplementary}. It is found that CuCl$_6$ octahedra are more distorted compared to the CuBr$_6$ one. In the basal plane, the Cu-$X$ bond distance for the Cl compound varies from 2.234 to 2.365~\AA~while the longest apical bond distances are in the range of 3.014 to 3.022~\AA. Similarly, for the Br compound the bond distances in the basal plane varies from 2.39~\AA~to 2.43~\AA~while the apical bond distances are in the range of 3.163 to 3.17~\AA. 

Each structural dimer unit of Cu$_2X_{10}$ is formed by edge sharing of two inequivalent Cu$X_6$ octahedra at the basal plane [see Fig.~\ref{Fig2}(b)]. Along the $b$-direction, the dimers are arranged parallel to each other and are interconnected via the edge sharing of Cu$X_6$ octahedra (between the apical and basal halide atoms) [see Fig.~\ref{Fig2}(c)]. In addition to the Cu-Cu intradimer distance $d_1$, if interdimer distance $d_3$ is taken into account, the spin-lattice would behave like a zig-zag two-leg ladder structure with $d_1$ and $d_3$ representing the rungs and legs of the ladder, respectively. The diagonal distance $d_2$ which is slightly less than $d_3$ makes the spin-lattice more intricate [see Fig.~\ref{Fig2}(d)]. These ladders are well apart and the organic cations reside in the interstitial space surrounding the ladders. Furthermore, the dimers from each ladder are aligned nearly perpendicular to the dimers of the neighbouring ladders [see Fig.~\ref{Fig2}(e)]. The value of bond distances $d_1$, $d_2$, and $d_3$ and the corresponding angles $\angle$Cu-$X$-Cu that favour the interaction paths $J_0$, $J^{\prime}$, and $J^{\prime \prime}$, respectively are tabulated in Table~\ref{bondlength}.
\begin{table*}
\setlength{\tabcolsep}{0.25 cm}
\caption{Values of bond lengths $d_1$, $d_2$, and $d_3$ as shown in Fig.~\ref{Fig2}(d) and angles $\angle$Cu-$X$-Cu as shown in Fig.~\ref{Fig2}(b) and (c) for both the compounds corresponding to the exchange pathways $J_0$, $J^{\prime}$, and $J^{\prime\prime}$.}
	\label{bondlength}
\begin{tabular}{*{9}{c} }
    \hline
   &\multicolumn{2}{c}{{$J_0$}}&&\multicolumn{2}{c}{{$J^{\prime}$}}
                    &&\multicolumn{2}{c}{{$J^{\prime\prime}$}} \\
\cline{2-3}\cline{5-6}\cline{8-9}
Compound & $d_1$ (\AA) & Angle (deg) && $d_2$ (\AA) & Angle (deg) && $d_3$ (\AA) & Angle (deg) \\
		\hline
 (C$_4$H$_{14}$N$_2$)Cu$_2$Cl$_6$ & 3.4730 & $\angle$Cu(2)-Cl(3)-Cu(1) && 3.8809 & $\angle$Cu(2)-Cl(3)-Cu(2)&& 3.933 &  
 $\angle$Cu(1)-Cl(3)-Cu(2) \\ 
		&&= 95.03 &&& = 91.95 &&& = 93.20\\
	&& $\angle$Cu(2)-Cl(4)-Cu(1) &&&&&& $\angle$Cu(1)-Cl(5)-Cu(2)  \\
		&& = 98.84 &&&&&& = 95.72 \\
       
(C$_4$H$_{14}$N$_2$)Cu$_2$Br$_6$ & 3.650 & $\angle$Cu(2)-Br(3)-Cu(1) && 4.070 & $\angle$Cu(2)-Br(4)-Cu(2) && 4.121 & $\angle$Cu(2)-Br(4)-Cu(1)\\
	&&= 97.61 &&& = 91.7 &&& = 92.77\\
	&&$\angle$Cu(2)-Br(4)-Cu(1) &&&&&& $\angle$Cu(2)-Br(5)-Cu(1)\\
	&&= 94.73 &&&&&& = 94.95\\  

  \hline  \hline
\end{tabular}
\end{table*} 


\subsection{Magnetization} 
\begin{figure*}
	\includegraphics[width = \linewidth]{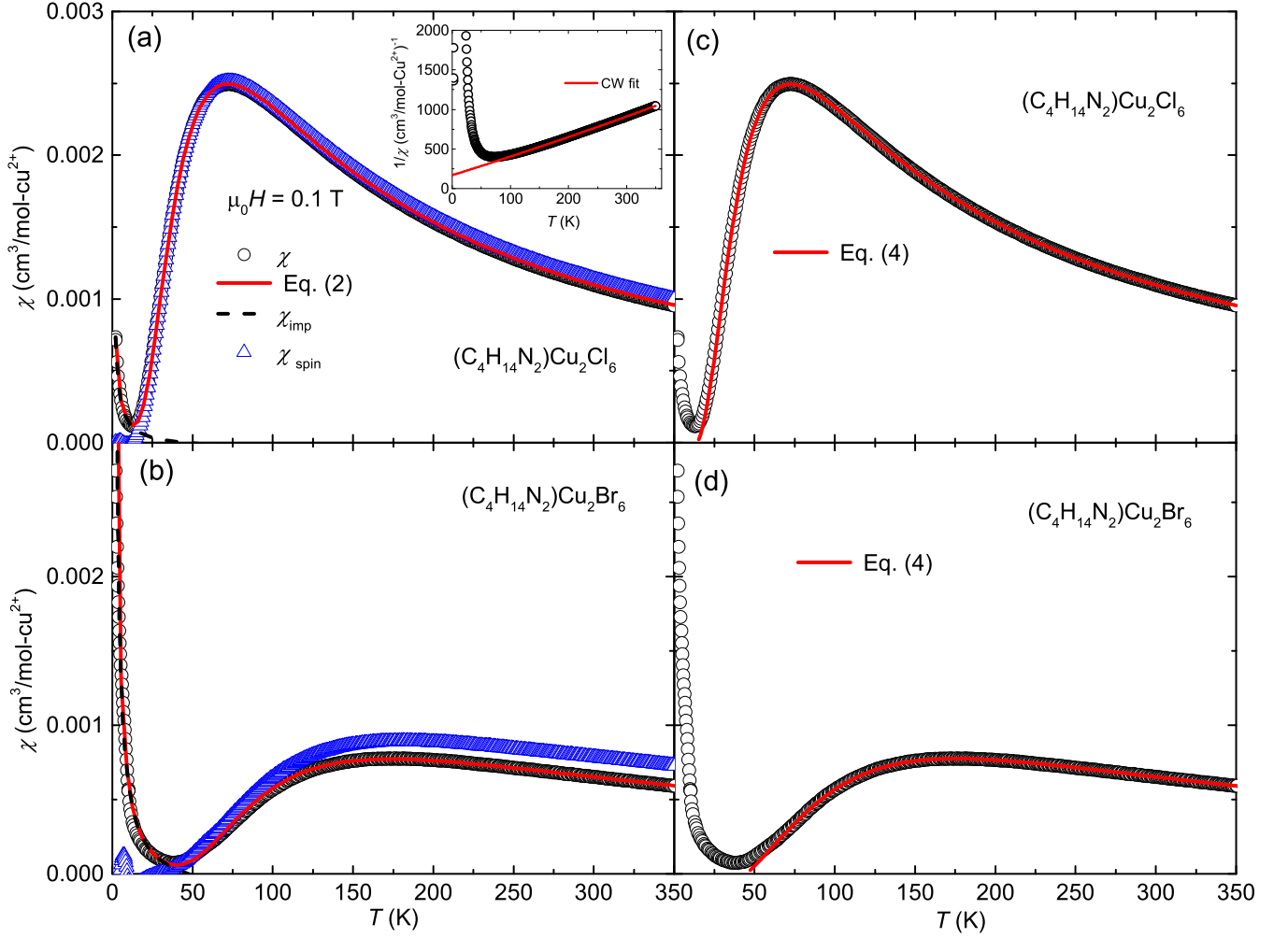}
	\caption{$\chi$ vs $T$ in an applied field of $\mu_{\rm 0}H = 0.1$~T for (a) (C$_4$H$_{14}$N$_2$)Cu$_2$Cl$_6$ and (b) (C$_4$H$_{14}$N$_2$)Cu$_2$Br$_6$. The solid line is the fit of the data using the spin-$\frac{1}{2}$ interacting-dimer model [Eq.~\eqref{chi_dimer}]. Inset: $1/\chi$ vs $T$ and the solid line is the CW fit for (C$_4$H$_{14}$N$_2$)Cu$_2$Cl$_6$. $\chi$ vs $T$ with isolated two-leg ladder model fit [Eq.~\eqref{chi_ladder}] for (c) (C$_4$H$_{14}$N$_2$)Cu$_2$Cl$_6$ and (d) (C$_4$H$_{14}$N$_2$)Cu$_2$Br$_6$.}
	\label{Fig3}
\end{figure*}

Figure~\ref{Fig3} presents the temperature dependent magnetic susceptibility ($\chi \equiv M/H$) of (C$_4$H$_{14}$N$_2$)Cu$_2X_6$ ($X=$ Cl, Br) measured in an applied field of $\mu_{0}H = 0.1$~T. Upon cooling, $\chi(T)$ of both compounds shows a Curie-Weiss (CW) increase in the high temperature regime and passes through a broad maxima at around $T_{\chi}^{\rm max}\simeq 72.5$ and 171.8~K, respectively, indicating the development of short-range antiferromagnetic (AFM) correlations. Below $T_{\chi}^{\rm max}$, $\chi(T)$ of both compounds falls rapidly, which is a primary indication of the opening of a spin-gap. At low temperatures, below 12.5~K (Cl) and 37.9~K (Br), $\chi(T)$ of these systems increases due to the presence of small fraction of extrinsic paramagnetic impurities or lattice imperfections in the samples. There is no signature of magnetic long-range order (LRO) down to 2~K for both compounds. 

$\chi(T)$ at high-temperatures was fitted by the sum of CW law and the temperature-independent susceptibility ($\chi_0$) 
\begin{equation}\label{CW}
\chi(T) = \chi_0 + \frac{C}{T - \theta_{\rm CW}}.
\end{equation}
Here, $C$ and $\theta_{\rm CW}$ are the Curie constant and CW temperature, respectively. The inverse susceptibility [1/$\chi(T)$] of (C$_4$H$_{14}$N$_2$)Cu$_2$Cl$_6$ was fitted above 200~K by Eq.~\eqref{CW} yielding $\chi_0\simeq-1.08\times 10^{-4}$~cm$^3$/mol-Cu$^{2+}$, $C\simeq0.45$~cm$^3$.K/mol-Cu$^{2+}$, and $\theta_{\rm CW}\simeq-75$~K. The negative value of $\theta_{\rm CW}$ indicates dominant AFM interaction between the Cu$^{2+}$ ions. The effective magnetic moment estimated from the $C$ value to be $\mu_{\rm eff}=(3k_{\rm B}C/N_{\rm A}\mu_{\rm B}^{2})^{\frac{1}{2}}\simeq 1.89$~$\mu_{\rm B}/$Cu$^{2+}$ (where $k_{\rm B}$ is the Boltzmann constant, $N_{\rm A}$ is the Avogadro’s number, and $\mu_{\rm B}$ is the Bohr magneton). This value of $\mu_{\rm eff}$ is slightly higher than the free ion value of $\mu_{\rm eff}$ ($1.73$~$\mu_{\rm B}$) for Cu$^{2+}$ with spin-$\frac{1}{2}$ and $g=2$. The $\mu_{\rm eff}$ value corresponds to $g\simeq 2.18$, typically observed for Cu$^{2+}$ based systems~\cite{Nath014407,Jin174423}. As the exchange coupling for (C$_4$H$_{14}$N$_2$)Cu$_2$Br$_6$ is relatively large and our $\chi(T)$ measurements are restricted to 350~K only, it was not possible to fit the data using Eq.~\eqref{CW}. This is because, CW fit requires data at very large temperature $T> \theta_{\rm CW}$~\cite{Somesh104422} and measurement above 350~K was not possible since both the compounds melt at around 420~K (see SM~\cite{Supplementary}).
 
To understand the exchange network, $\chi(T)$ data over the whole temperature range were fitted by the equation
\begin{equation}
\label{chi_dimer}
\chi(T) = \chi_0 + \frac{C_{\rm imp}}{T - \theta_{\rm imp}}+\chi_{\rm spin}(T).
\end{equation}
In the second term, $C_{\rm imp}$ is the Curie constant of the impurity spins and $\theta_{\rm imp}$ is the effective interaction strength between the impurity spins. This term takes care of the low-temperature Curie tail in $\chi(T)$. $\chi_{\rm spin}(T)$ is the spin susceptibility of the spin-$\frac{1}{2}$ interacting dimer model which has the form~\cite{Arjun174421}
\begin{eqnarray}
\label{chi_spin}
\chi_{\rm spin}(T)=\frac{N_{\rm A}g^{2}\mu_{\rm B}^{2}}{k_{\rm B}T}\frac{1}{\left[3+e^{\left(\frac{J_0}{k_{\rm B}T}\right)}+\frac{z^{\prime}J^{\prime}}{k_{\rm B}T}\right]}.
\end{eqnarray}
Here, $J_0$ and $J^{\prime}$ are the intra- and average inter-dimer interactions, respectively. In this expression, a mean-field approximation is used to introduce $J^{\prime}$ in the isolated dimer model~\cite{Thamban255801}. Here, $z^{\prime} = 2$ represents the number of neighbouring dimers coupled with one dimer through $J^{\prime}$. As shown in Fig.~\ref{Fig3}, Eq.~\eqref{chi_dimer} reproduces our experimental data very well resulting the parameters ($\chi_0\simeq-4.52\times10^{-5}$~cm$^3$/mol-Cu$^{2+}$, $C_{\rm imp}\simeq0.002$~cm$^3$.K/mol-Cu$^{2+}$, $\theta_{\rm imp}\simeq-0.64$~K, $g=2.05$, $J_0/k_{\rm B}\simeq116.7$~K, and $J^{\prime}/k_{\rm B}\simeq25$~K) and ($\chi_0\simeq-1.55\times10^{-4}$~cm$^3$/mol-Cu$^{2+}$, $C_{\rm imp}\simeq0.00725$~cm$^3$.K/mol-Cu$^{2+}$, $\theta_{\rm imp}\simeq0.53$~K, $g=2.12$, $J_0/k_{\rm B}\simeq288.8$~K, and $J^{\prime}/k_{\rm B}\simeq235$~K) for (C$_4$H$_{14}$N$_2$)Cu$_2$Cl$_6$ and (C$_4$H$_{14}$N$_2$)Cu$_2$Br$_6$, respectively.
The obtained values of $C_{\rm imp}$ correspond to $\sim 0.53$~\% and $\sim 1.9$~\% of the paramagnetic impurity spins, respectively, assuming spin-$\frac{1}{2}$. In order to emphasize the gapped behavior, $\chi_0 + \frac{C_{\rm imp}}{T - \theta_{\rm imp}}$ was subtracted from the $\chi(T)$ data. The resulting intrinsic $\chi_{\rm spin}(T)$ indeed decays exponentially towards zero at low temperatures [see Fig.~\ref{Fig3}(a) and (b)] further establishing a singlet ground state.

As demonstrated in Fig.~\ref{Fig2}(d), there is equal possibility for the parallel dimers to interact along the legs of the ladder. Therefore, we fitted the $\chi(T)$ data by the high-$T$ series expansion (HTSE) of strong rung ladder model as
\begin{equation}
\label{chi_ladder}
\chi(T) = \chi_0 + \chi_{\rm spin}(T).
\end{equation}
Here, $\chi_{\rm spin}$ is the expression of HTSE for spin-1/2 two-leg ladder with strong rung coupling [i.e. Eq.~(47) in Ref.~\cite{Johnston2000}]. This expression is valid in high temperatures and for $0 \leq J^{\prime \prime}/J_0 \leq 1$. This expression predicts more accurate results for $J^{\prime \prime}/J_0 < 0.6$. Our fit in the high-$T$ regime $T > 0.2 J_0$ (i.e. 25~K for Cl and 68~K for Br) reproduces the experimental data very well yielding ($\chi_0 \simeq -4.99 \times 10^{-5}$~cm$^3$/mol, $g \simeq 2.06$, $J_0/k_{\rm B} \simeq115$~K, and $J^{\prime \prime}/k_{\rm B}\simeq 22.9$~K) and ($\chi_0\simeq-1.108\times 10^{-4}$~cm$^3$/mol, $g\simeq 2.06$, $J_0/k_{\rm B} \simeq 270.8$~K, and $J^{\prime\prime}/k_{\rm B} \simeq 99.6$~K) for Cl and Br compounds, respectively [see Fig.~\ref{Fig3}(c) and (d)].

The zero-field spin-gap ($\Delta_0$) for a strong rung coupled two-leg ladder can be estimated as~\cite{Johnston2000}
\begin{equation}
	\label{ladder energy gap}
	\Delta_0 = J_0-J^{\prime\prime}+\frac{{{J^{\prime\prime}}^2}}{2J_0}+\frac{{{J^{\prime\prime}}^3}}{4J_0^2}-\frac{{{J^{\prime\prime}}^4}}{8J_0^3}+\mathcal O ({{J^{\prime\prime}}^5)}.
\end{equation}
Using the appropriate values of $J_0$ and $J^{\prime\prime}$, $\Delta_0$ is calculated to be 94.7~K and 192.3~K for the Cl and Br compounds, respectively. The values of spin-gap correspond to the critical field of gap closing $H_{\rm C1} \simeq 68.4$~T and 146.4~T for the Cl and Br compounds, respectively. Similarly, the saturation field ($H_{\rm C2}$) at which one can achieve the fully polarized state is calculated to be $H_{\rm C2} = (J_0+2J^{\prime \prime})/g\mu_{\rm B} \simeq 115$~T and 358~T for the Cl and Br compounds, respectively~\cite{Landee100402}.

\begin{figure}
	\includegraphics {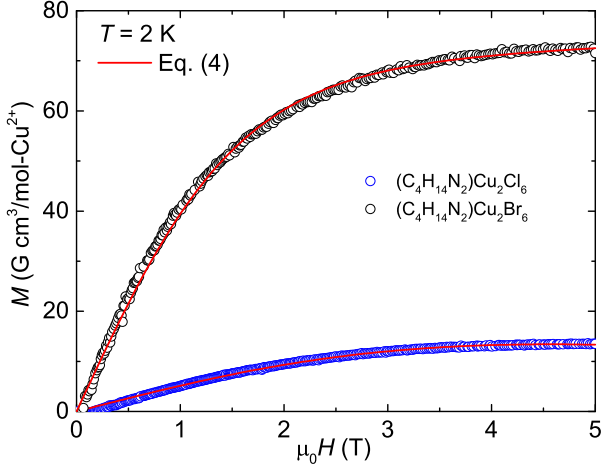}
	\caption {Magnetization isotherms ($M$ vs $H$) at $T=2$~K measured up to 5~T. Solid lines are the Brillouin fit using Eq.~\eqref{MH imp}.} 
	 \label{Fig4}
\end{figure}
Magnetic isotherm ($M$ vs $H$) measured at $T=2$~K is shown in Fig.~\ref{Fig4}. For both the compounds, it shows a typical paramagnetic behavior up to 5~T. The $\chi(T)$ analysis suggests that $\chi_{\rm spin}(T)$ approaches zero value below $\sim13$~K and $\sim37$~K for (C$_4$H$_{14}$N$_2$)Cu$_2$Cl$_6$ and (C$_4$H$_{14}$N$_2$)Cu$_2$Br$_6$, respectively and the low-$T$ upturn in $\chi(T)$ is entirely due to extrinsic contributions. As one expects zero magnetization in the gapped state, the observed $M$ vs $H$ behaviour can be attributed completely due to the paramagnetic impurity spins. Hence, one can estimate this extrinsic paramagnetic contribution accurately by fitting the data to~\cite{Nath174513}
\begin{equation}
	\label{MH imp}
	M(H)=\chi H + f_{\rm imp}N_{\rm A}g_{\rm imp}\mu_{\rm B}S_{\rm imp}Bs_{\rm imp}(x).
\end{equation}
In the above equation, $\chi$ is the intrinsic susceptibility, $f_{\rm imp}$ is the molar fraction of the impurities, $g_{\rm imp}$ is the impurity $g$-factor, $S_{\rm imp}$ is the impurity spin, $Bs_{\rm imp}(x)$ is the Brillouin function, and $x= g_{\rm imp}\mu_{\rm B}S_{\rm imp}H/[k_{\rm B}(T-\theta_{\rm imp})]$. We assumed the impurity spin $S_{\rm imp} = 1/2$ and the Brillouin function reduces to $Bs_{\rm imp}(x)= \text{tanh}(x)$~\cite{Kittel1986}. Our fitted results upon fixing $g=2$ are ($f_{\rm imp} \simeq 0.0048$~mol\% and $\theta_{\rm imp} \simeq -0.44$~K) and ($f_{\rm imp} \simeq 0.0122$~mol\% and $\theta_{\rm imp} \simeq 0.95$~K) for the Cl and Br compounds, respectively. 
The obtained values of $f_{\rm imp}$ correspond to $\sim 0.48$~\% and $\sim 1.2 $~\% of spin-$\frac{1}{2}$ paramagnetic impurity spins for the Cl and Br compounds, respectively which are consistent with the $\chi(T)$ analysis.

\subsection{Quantum Entanglement}
\begin{figure}
	\includegraphics[scale=1.1]{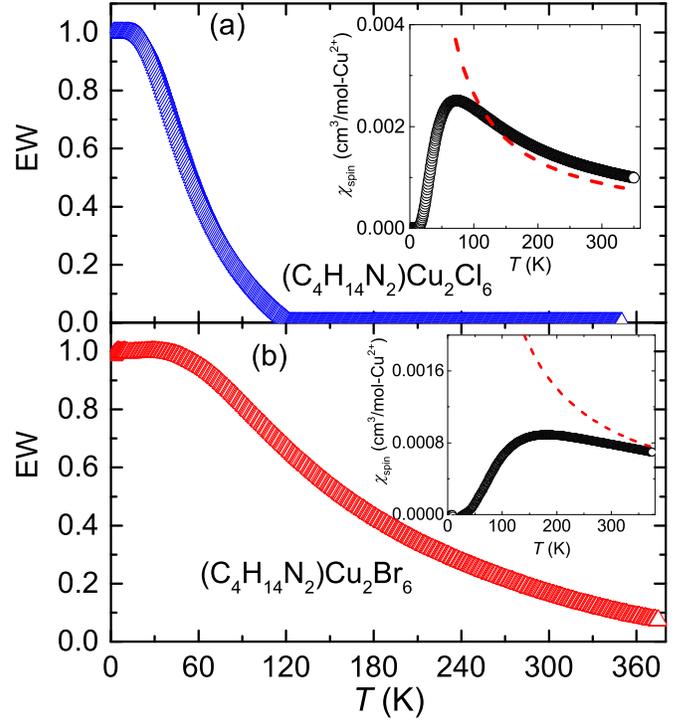}
	\caption{Entanglement witness (EW) as a function of temperature for (a) (C$_4$H$_{14}$N$_2$)Cu$_2$Cl$_6$ and (b) (C$_4$H$_{14}$N$_2$)Cu$_2$Br$_6$. Inset: $\chi_{\rm spin}$ vs $T$ along with the entanglement boundary (dashed line).}	
	\label{Fig5}
\end{figure}
Existence of entanglement in an AFM spin system can be measured by a quantity called entanglement witness (EW). In a magnetic system with singlet ground state, the two spins are strongly entangled and one can extract EW from the macroscopic thermodynamic observable like $\chi_{\rm spin}$. For a spin-$1/2$ isotropic Heisenberg system, EW is related to $\chi_{\rm spin}$ as~\cite{Wiesniak258}
\begin{eqnarray}
	\label{EW}
	{\rm EW}=1-\frac{6k_{\rm B}T\chi_{\rm spin}}{N_{\rm A}g^{2}\mu_{\rm B}^{2}}.
\end{eqnarray}
Figure~\ref{Fig5} depicts the temperature variation of EW calculated using Eq.~\eqref{EW}. EW of both compounds reaches a maximum value of 1 in the low temperature region where $\chi_{\rm spin}$ is zero. It decreases with increasing temperature. For (C$_4$H$_{14}$N$_2$)Cu$_2$Cl$_6$, EW approaches zero at around 120~K, whereas for (C$_4$H$_{14}$N$_2$)Cu$_2$Br$_6$, it remains non-zero even at 370~K. 
For an entangled state, EW should have a finite value ($> 0$). The dashed line [Eq.~\eqref{EW} taking EW = 0] in the insets of Fig.~\ref{Fig5}(a) and (b) represents the boundary of the entangled state, which is plotted along with $\chi_{\rm spin}(T)$.
The point of intersection with $\chi_{\rm spin}$ defines the upper temperature limit of the entangled state.
The dashed curve intersects the $\chi_{\rm spin}$ data of (C$_4$H$_{14}$N$_2$)Cu$_2$Cl$_6$ at around 120~K, which is consistent with the above analysis. On the other hand, (C$_4$H$_{14}$N$_2$)Cu$_2$Br$_6$ demonstrates that entanglement persists even beyond 370~K. Experimentally, quantum entanglement is realized in several spin-1/2 Heisenberg AFM dimers and spin-chain systems such as Cu(NO$_3)_2${\textperiodcentered}2.5H$_2$O~\cite{Singh951}, Na$_2$Cu$_5$Si$_4$O$_{\rm 14}$~\cite{Souza104402}, NH$_4$CuPO$_4${\textperiodcentered}H$_2$O~\cite{Chakraborty034909}, and Cu(tz)$_{2}$Cl$_{2}$~\cite{Chakraborty2967}. However, all these compounds are entangled at very low temperatures except Na$_2$Cu$_5$Si$_4$O$_{\rm 14}$. In this context, (C$_4$H$_{14}$N$_2$)Cu$_2$Cl$_6$ and (C$_4$H$_{14}$N$_2$)Cu$_2$Br$_6$ are two promising compounds where entanglement perseveres up to much higher temperatures compared to the above mentioned compounds.

\subsection{Theoretical Calculations} 
\begin{figure}
	\includegraphics[scale=0.7] {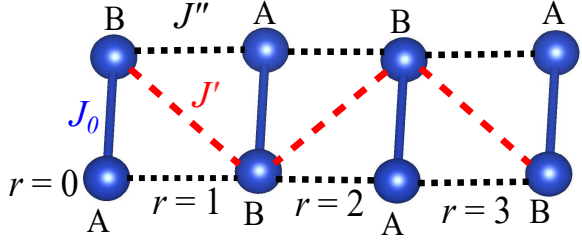}
	\caption {Lattice structure: A and B are two sub-lattices of the ladder structure. $J_0$ and $J^{\prime\prime}$ are rung and leg couplings, respectively while $J^{\prime}$ is the coupling along the diagonal. $r$ is the site index along the lower leg of the system.}
	 \label{Fig6}
\end{figure}

A frustrated two-leg ladder is pictorially shown in Fig.~\ref{Fig6} where each solid sphere represents a Cu$^{2+}$ site with spin-1/2. The two in-equivalent Cu sites are denoted as A and B. An isotropic spin-1/2 Heisenberg model Hamiltonian on this lattice can be written as:
\begin{equation}\label{Hamil_id}
\begin{aligned}
\mathcal{H}=\sum_{i=1}^{N/2}J_0 \vec{S}_{i,A}.\vec{S}_{i,B}+\sum_{i=1}^{N/2}J^\prime \vec{S}_{i,B}.\vec{S}_{i+1,B}+\\
\sum_{i=1}^{N/2}J^{\prime\prime} (\vec{S}_{i,B}.\vec{S}_{i+1,A} + \vec{S}_{i,A}.\vec{S}_{i+1,B})\\
-H\sum_{i=1}^{N/2}(S_{i,A}^z+S_{i,B}^z),
\end{aligned}
\end{equation}
where $\vec{S}_{i,A}$ and $\vec{S}_{i,B}$ indicate the spin vectors for the sub-lattices A and B, respectively of the $i^{th}$ unit cell. In Eq.~\eqref{Hamil_id}, the first term represents the intra-dimer coupling $J_0$ while the second and third terms represent the inter-dimer coupling $J^\prime$ and $J^{\prime\prime}$ between the neighboring dimers along the diagonal and the leg, respectively. The coupling $J^\prime$ is always between two same Cu sites while $J^{\prime\prime}$ is between two in-equivalent Cu sites. The last term of the Hamiltonian represents an externally applied axial magnetic field $H$, and the energy scale is set in terms of $J_0$.


In order to have a further insight about the ground state properties, correlation function, $C(r)= \expval{S_i.S_{i+r}}{\Psi}$ vs distance $r$ of the lattice sites and magnetization $M=\expval{\sum_{i=1}^{N}s_i^z}{\psi}$ vs magnetic field ($H$) calculations are performed up to $N=24$ sites by considering frustrated two leg-ladder model. The magnetization $M(T,H)$ and magnetic susceptibility $\chi(T,H)$ of these systems can be calculated using the full spectrum and the partition function $Z(T,H)$ can be written as:
\begin{equation}\label{CW1}
Z(T,H)=\sum_{s^z=-N/2}^{N/2}\sum_{n_{s^z}} e^{-\beta(E_{n_{s^z}}-Hs^z)}.
\end{equation}
Here, $\beta=\frac{1}{k_{\rm B}T}$ and $s^z$ is the $z$-component of total spin which varies from $-N/2$ to $N/2$ where total number of sites in the system are $N$. $E_{n_{s^z}}$ is the energy for $n_{s^z}$ state. 
The magnetization can be defined as
\begin{equation}\label{CW2}
M(T,H)=\frac{1}{N}\sum_{s^z=-N/2}^{N/2}\sum_{n_{s^z}}s^ze^{-\beta(E_{n_{s^z}}-Hs^z)}.
\end{equation}
$\chi(T,H)$ can be written in terms of the magnetic fluctuation as
\begin{equation}\label{CW3}
\chi(T,H)=\frac{\beta}{N}[\langle M^2 \rangle - {\langle M \rangle}^2].
\end{equation}

\begin{figure}
	\includegraphics[width = \linewidth] {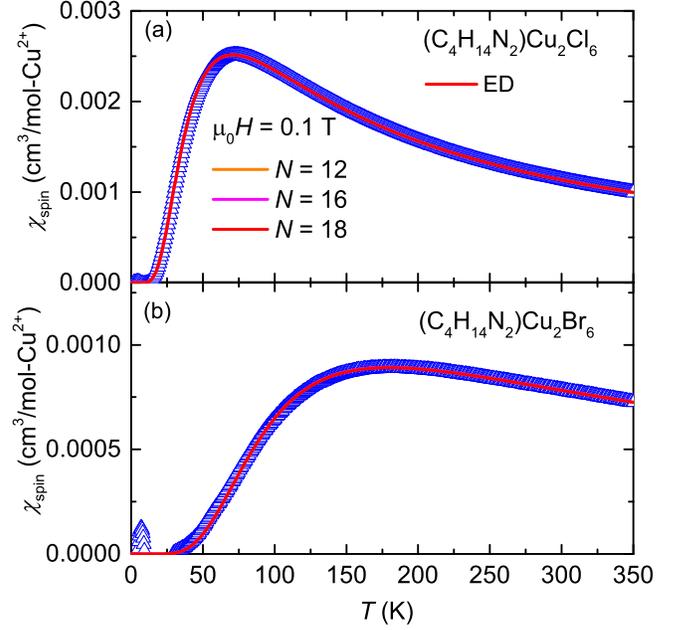}
	\caption {The experimental $\chi_{\rm spin}$ data (symbols) for (a) (C$_4$H$_{14}$N$_2$)Cu$_2$Cl$_6$ and (b) 
 (C$_4$H$_{14}$N$_2$)Cu$_2$Br$_6$. The fits using ED simulation for the frustrated two-leg ladder are shown as solid lines for system sizes $N = 12$, 16, and 18.} 
	 \label{Fig7}
\end{figure}
First, we calculated the magnetic susceptibility $\chi(T)$ by taking frustrated two-leg ladder model into account for both materials. As the experimental $\chi(T)$ contains a large low temperature upturn (see Fig.~\ref{Fig3}), to fit the experimental data, we used the spin susceptibility $\chi_{\rm spin}$ originating from the interacting dimer model fit in Fig.~\ref{Fig3}. In order to show the finite-size effect, $\chi_{\rm spin}$ is calculated for three different system sizes, $N=12$, 16, and 18. Clearly, there is no visible finite-size effect for both compounds. From the best fit of the experimental $\chi_{\rm spin}$ data in Fig.~\ref{Fig7}, the obtained parameters are ($g=2.06$, $J_0/k_{\rm B}=116$~K, $J^\prime/k_{\rm B}= 23.2$~K, and $J^{\prime\prime}/k_{\rm B}= 18.56$~K) for (C$_4$H$_{14}$N$_2$)Cu$_2$Cl$_6$ and ($g=2.06$, $J_0/k_{\rm B}=300$~K, $J^\prime/k_{\rm B} = 90$~K, and $J^{\prime\prime}/k_{\rm B} = 105$~K) for (C$_4$H$_{14}$N$_2$)Cu$_2$Br$_6$, respectively.

\begin{figure}
	\includegraphics[scale=1.1] {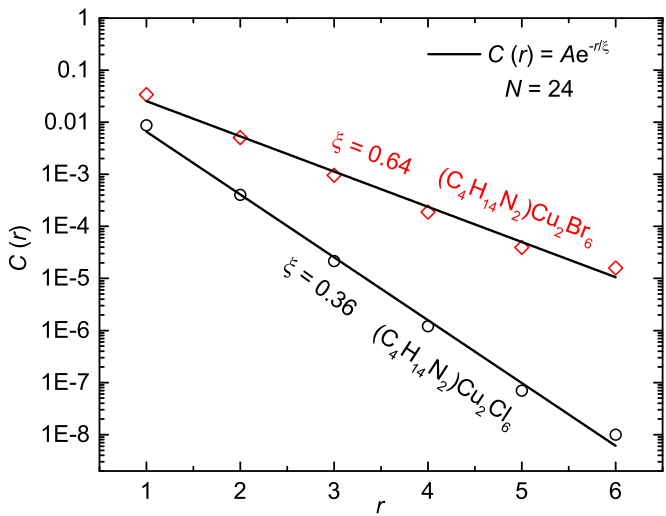}
	\caption {The correlation function $C(r)$ vs $r$ calculated for the frustrated two-leg ladder model along the lower leg by taking $r=0$ as reference site (see Fig.~\ref{Fig6}) for (C$_4$H$_{14}$N$_2$)Cu$_2$Cl$_6$ and (C$_4$H$_{14}$N$_2$)Cu$_2$Br$_6$. Solid lines are the exponential fits.}
	 \label{Fig8}
\end{figure}
We also calculated the correlation function $C(r)$ for the frustrated two-leg ladder model for both compounds along the A-B-A-B-... sites [see Fig.~\ref{Fig6}], where $r$ is the distance. We notice that $C(r)$ decays exponentially along the leg (A-B-A-B-...) and it can be fitted well with an exponential function $Ae^{-r/\xi}$ where $A$ is a constant. From the fit in Fig.~\ref{Fig8}, the correlation length $\xi$ is estimated to be $\sim 0.36$ and $\sim 0.64$ for (C$_4$H$_{14}$N$_2$)Cu$_2$Cl$_6$ and (C$_4$H$_{14}$N$_2$)Cu$_2$Br$_6$, respectively, which are still less than the spacing between nearest-neighbour Cu$^{2+}$ ions along the leg. Surprisingly, for (C$_4$H$_{14}$N$_2$)Cu$_2$Br$_6$, despite large $J^{\prime}/J_0$ and $J^{\prime\prime}/J_0$ ratios ($\sim 0.30$ and $0.35$), $\xi$ is less than the lattice spacing and it is still behaving like weakly coupled dimers.

\begin{figure}
	\includegraphics[scale=1] {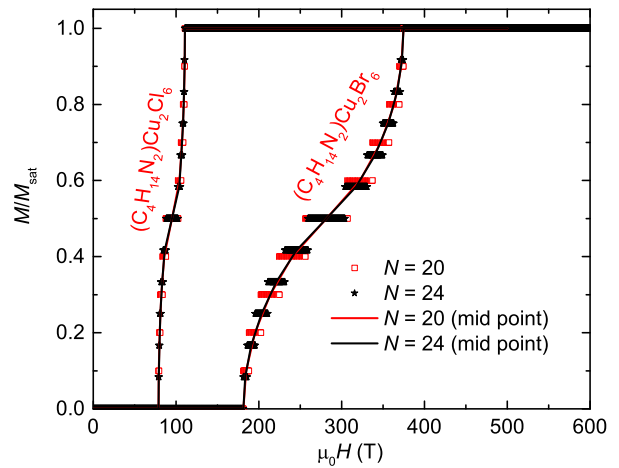}
	\caption {The calculated $M$ vs $H$ curves for (C$_4$H$_{14}$N$_2$)Cu$_2$Cl$_6$ and (C$_4$H$_{14}$N$_2$)Cu$_2$Br$_6$ with system sizes, $N = 20$ and 24 using the frustrated two-leg ladder model. Symbols and solid lines represent the plateaus and connector of the mid point of plateaus, respectively.}
	 \label{Fig9}
\end{figure}
The spin-gap and saturation field are two important quantities to understand the magnetic properties of a gapped spin system. Due to large energy scale of the exchange couplings, the critical fields are too high and are not accessible experimentally. Hence, to extract the value of critical fields (critical field of gap closing $H_{\rm C1}$ and saturation field $H_{\rm C2}$) and to understand the nature of magnetic isotherms, we have simulated the $M$ vs $H$ curve at zero temperature for both compounds for $N = 20$ and 24 sites (see Fig.~\ref{Fig9}). The critical fields
are found to be independent of the system size for both the compounds. In the intermediate field regime, the Cl compound has very weak finite size effect while the Br compound exhibits a plateau-like behavior and should have a continuous $M$ vs $H$ curve in the thermodynamic limit ($N \to \infty$). Finite size effects can be avoided by considering the value of $M$ at the midpoint of each plateau and drawing a continuous line through these points for $N=20$ and 24 which show almost no difference. The value of $H_{\rm C1}$ is found to be 78.9~T and 181.2~T for the Cl and Br compounds, respectively. 
This suggests that such a robust singlet ground state cannot be perturbed by the experimentally available magnetic fields.
The saturation magnetic field or the field required to obtain the fully polarized state is found to be $H_{\rm C2} = 110.7$~T and 374.2~T for the Cl and Br compounds, respectively.
These values of $H_{\rm C1}$ and $H_{\rm C2}$ are slightly larger than the ones predicted from the analysis using two-leg ladder model, as the two-leg ladder model doesnot take into account the diagonal interaction.

\section{Discussion and Summary}
We investigated the structural and magnetic properties of two copper halides. Though, experimental $\chi(T)$ data are fitted well by both spin-$1/2$ interacting dimer and two-leg ladder models, a more accurate description is obtained from the ED calculations. The ED calculation results assuming a two-leg ladder model reproduce the spin susceptibility more precisely with a strong rung coupling $J_0/k_{\rm B} = 116$~K and 300~K, a weak but significant leg coupling $J^{\prime \prime}/k_{\rm B} = 18.6$~K and 105~K, and another weak diagonal coupling $J^{\prime}/k_{\rm B} = 23.2$~K and 90~K for (C$_4$H$_{14}$N$_2$)Cu$_2$Cl$_6$ and (C$_4$H$_{14}$N$_2$)Cu$_2$Br$_6$, respectively. Depending on the hierarchy of the exchange couplings, two-leg spin-1/2 ladders may evince
remarkable critical behaviour at low temperatures due low dimensionality and reduced spin value. For instance, though, TLL physics is envisaged in two-leg ladders, a qualitative difference is delineated between strong-leg and strong-rung ladders. Theory predicts attractive fermionic interaction in strong-leg spin-1/2 ladders [e.g. (C$_7$H$_{10}$N)$_2$CuBr$_4$] and repulsive fermionic interaction in strong-rung spin-1/2 ladders [e.g. (C$_5$H$_{12}$N)$_2$CuBr$_4$], respectively in the gapless TLL critical state~\cite{Jeong106402,Jeong106404}. Experimental verification of such predictions are strongly hindered due to the unavailability of model compounds. Moreover, unlike strong-leg coupled two-leg ladders, the number of strong-rung coupled two-leg ladder compounds are limited till date~\cite{Watson5168,Landee100402}. Hence, the compounds under investigation are an important class of compounds in this direction. Further, as depicted in Fig.~\ref{Fig6}, the AFM $J^{\prime}$ along the diagonal induces magnetic frustration in the ladder, making it an unique spin-lattice of frustrated two-leg ladder in both our compounds~\cite{Tsirlin144426}.

According to the Goodenough-Kanamori-Anderson (GKA) rule, the nature and strength of the superexchange in magnetic insulators depend very much on the Cu–$X$–Cu bridging angle and the extent of overlap between the Cu~$3d$ and $X$~$3p/4p$ orbitals of the ligands ($X$ = Cl, Br). Strong AFM exchange interaction is favoured for $\angle Cu-X-Cu > 95^0$~\cite{Goodenough564,*Kanamori87,*Anderson99}.
As discussed earlier, the Cu$X_6$ octahedra are distorted due to Jahn-Teller effect and the Cu-$X$ bond is elongated along the apical direction. In an octahedral coordination, 3$d_{x^2-y^2}$ orbital of Cu$^{2+}$ lie in the basal plane and contains the unpaired electron. In both the compounds, the intra-dimer (or rung) coupling ($J_0$) arises due to the overlapping of 3$d_{x^2-y^2}$ orbitals of two Cu$^{2+}$ ions via the quasi-orthogonal $p$ orbitals of the halide ligands in the basal plane.
On the other hand, the diagonal ($J^{\prime}$) and leg ($J^{\prime \prime}$) interactions emerge from to the overlapping of 3$d_{z^2}$ orbitals of one Cu$^{2+}$ with $3d_{x^2-y^2}$ orbitals of another Cu$^{2+}$ from the neighbouring dimers via the $p$ orbitals of the apical halide ions of the distorted Cu$X_6$ octahedra. Furthermore, the intra-dimer distance $d_1$ is found to be smaller than $d_2$ and $d_3$ and the angle $\angle Cu-X-Cu$ along $d_1$ is more than $95^0$ as well as more than the angles along $d_2$ and $d_3$ (see Table~\ref{bondlength}). Thus, the presence of unpaired electron in the 3$d_{x^2-y^2}$ orbital, shorter bond distance, and larger bong angle are the obvious reasons for much stronger AFM coupling for $J_0$ as compared to $J^{\prime}$ and $J^{\prime \prime}$.
Due to extended Cu-$X$-$X$-Cu paths, the orthogonal dimers of the neighboring ladders are very weakly connected in the $ac$-plane through the basal halide ions. The magnetic network of these compounds seems to be similar to the celebrated compounds $A$Cu$X_3$ ($A$ = Tl, NH$_4$, K)~\cite{Matsumoto054423}.

Despite the same crystal structure, the exchange couplings ($J_0$, $J^{\prime}$, and $J^{\prime \prime}$) of (C$_4$H$_{14}$N$_2$)Cu$_2$Br$_6$ are larger than (C$_4$H$_{14}$N$_2$)Cu$_2$Cl$_6$.
The Cu-Cu bond distance and $\angle$Cu-$X$-Cu bond angles (see Table~\ref{bondlength}) for both the compounds are nearly same and making a distinction is really difficult.
The superexchange mechanism between Cu$^{2+}$ ions is explained well for Cu$X_2$ ($X=$~F, Cl, and Br) in Ref.~[\onlinecite{Lebernegg155111}] in terms of the GKA rule~\cite{Goodenough564,*Kanamori87,*Anderson99}.
It is reported that the larger ligand size can amplify the orbital overlap mechanism, leading to larger exchange couplings. Since, Br$^{-}$ has a large ionic radius than Cl$^{-}$, one expects higher $J_0$, $J^{\prime}$, and $J^{\prime \prime}$ for (C$_4$H$_{14}$N$_2$)Cu$_2$Br$_6$ than (C$_4$H$_{14}$N$_2$)Cu$_2$Cl$_6$.
Similar results are reported for the alternating chain compound, (4,4$^{\prime}$-bipyridinium)Cu$_2$Cl$_{6-x}$Br$_{x}$, where the AFM couplings are increased monotonously with increasing bromide concentration~\cite{Willett2093}. Moreover, distortion in the Cu$X_6$ octahedra units also plays a decisive role. In case of CuCl$_6$, the distortion is more than CuBr$_6$ which possibly favours weaker interaction in (C$_4$H$_{14}$N$_2$)Cu$_2$Cl$_6$.

For an isolated dimer compound, the intra-dimer coupling represents the true spin-gap in zero-field and the critical field of gap closing ($H_{\rm C1}$) coincides with the critical field of saturation ($H_{\rm C2}$). The inter-dimer coupling not only reduces the value of spin-gap but also increases the spread between $H_{\rm C1}$ and $H_{\rm C2}$, allowing space for field induced quantum phases to occur~\cite{Giamarchi198,Ruegg62}. As shown in Fig.~\ref{Fig9}, these fields are $H_{\rm C1}\simeq 78.9$~T and $H_{\rm C2}\simeq 110.7$~T for the Cl compound and $H_{\rm C1} \simeq 181.2$~T and $H_{\rm C2} \simeq 374.2$~T for the Br compound. Thus, the large spacing between $H_{\rm C1}$ and $H_{\rm C2}$ implies significant inter-dimer coupling or leg coupling along the ladder for both the compounds, consistent with our $\chi(T)$ analysis.
Unfortunately, their exchange couplings are so large that the critical fields are not easily accessible experimentally. Nevertheless, these compounds have great potential if new materials can be designed with reduced exchange couplings by appropriately choosing the halide ions and ligands or introducing more distortion in the lattice. 
Another interesting aspect of these compounds is that the dimers from one ladder are nearly orthogonal to the dimers in the neighbouring ladders. Therefore, if the inter-dimer couplings can be weakened by tuning structural parameters then these compounds can be considered equivalent to the Shastry-Sutherland lattices Sr$_2$Cu(BO$_3$)$_2$ and BaNd$_2$ZnO$_5$~\cite{Kageyama3168,Ishii064418}.
Thus, synthesis of these compounds opens up a new route to tune the magnetic parameters and realize new compounds that would be relevant from the quantum magnetism point of view.

In summary, we present the details of single-crystal growth and magnetic properties of two interesting iso-structural quantum magnets (C$_4$H$_{14}$N$_2$)Cu$_2X_6$ ($X$= Cl, Br).
Both the compounds feature Cu$^{2+}$ two-leg ladders with a frustrated diagonal coupling. The dimers from each ladder are aligned orthogonal to the dimers from the adjacent ladders. The analysis of $\chi(T)$ and the subsequent theoretical calculations reveal that the dimers are strongly coupled leading to a drastic reduction in spin-gap from its isolated dimer value. Our exact diagonalization calculations assuming a frustrated two-leg ladder geometry, indeed, reproduce the experimental $\chi(T)$ with leading rung coupling $J_0/k_{\rm B} \simeq 116$~K and 300~K, weak leg coupling $J^{\prime}/k_{\rm B} \simeq 18.6$~K and 105~K, and weak frustrated diagonal coupling $J^{\prime \prime}/k_{\rm B} \simeq 23.2$~K and 90~K for the Cl and Br compounds, respectively. 
Despite same crystal symmetry, the relatively large exchange couplings in case of Br as compared to Cl compound is attributed to larger ionic size of Br and more diffused $p$-orbitals that facilitate stronger coupling between the Cu$^{2+}$ ions.
Further tuning of exchange couplings by appropriately choosing the halide ions or legands would make them model compounds for exploring field induced quantum phases.

\section{Acknowledgement}
SG and RN acknowledge SERB, India for financial support bearing sanction Grant No.~CRG/2022/000997. SG is supported by the Prime Minister’s Research Fellowship (PMRF) scheme, Government of India. MK thanks SERB for financial support through Grant Sanction No. CRG/2020/000754. S. Ghosh would like to express sincere gratitude to the DST-Inspire for financial support. DS acknowledges financial support from the Max Planck partner group and SERB, India (research grant No.~CRG/Z019/005144).

%

\newpage

\begin{titlepage}
	\centering
	{\Large\textbf{Crystal structure and magnetic properties of spin-$1/2$ frustrated two-leg ladder compounds (C$_4$H$_{14}$N$_2$)Cu$_2X_6$ ($X$= Cl and Br)}\par}
 Bottom of the page
	{\large \today\par}
\end{titlepage}

\begin{table*}
	\setlength{\tabcolsep}{0.2cm}
	\caption{The atomic coordinates ($x,y,z$) for (C$_4$H$_{14}$N$_2$)Cu$_2$Cl$_6$. $U_{\rm iso}$ is the isotropic atomic displacement parameters which is defined as one-third of the trace of the orthogonal $U_{\rm ij}$ tensor. The errors are from the least-square structure refinement. The positions of hydrogen atoms are fixed.}
	\label{atmic position_Cl}
	\begin{tabular}{ccccccc}
		\hline \hline
		Atomic sites & $x$ & $y$ & $z$ & $U_{\rm iso}$(\AA$^{2}$) & \\\hline
		Cu(1) & 0.07498(3)& 0.250000& 0.82571(3)& 0.02590(14)\\
		Cu(2) & -0.08156(3)& 0.250000& 1.00067(3)& 0.02298(14)\\
		C(1) & 0.1128(3)& 0.250000& 1.3219(3)& 0.0365(10)\\            
		C(3) & 0.2071(3)& 0.250000& 1.1846(2)& 0.0333(9)\\            
		C(4) &0.3007(3)& 0.250000& 1.1488(2)& 0.0326(9)\\          
		C(6) & 0.3923(3)& 0.250000& 1.0098(3)& 0.0436(11)\\                   
		N(1) & 0.20431(19)& 0.250000& 1.28540(19)& 0.0256(7)\\    
		N(2) &0.3016(2)& 0.250000& 1.04775(19) &0.0289(7)\\                
		Cl(1) & 0.22607(6)& 0.250000 &0.82506(6)& 0.0283(2) \\
		Cl(2) & 0.05884(7)& 0.250000& 0.67277(6)& 0.0358(3) \\      
		Cl(3) & 0.07383(5)& 0.250000& 0.98775(5)& 0.0228(2)\\
		Cl(4) &-0.07618(6)& 0.250000& 0.84431(6)& 0.0267(2)\\           
		Cl(5) &-0.07465(6)& 0.250000& 1.15362(6)& 0.0281(2) \\
		Cl(6) & -0.23268(6)& 0.250000& 0.99586(6)& 0.0276(2)\\          
		H(1A) &0.232782 &0.369578& 1.306130& 0.031\\  
		H(1B ) & 0.232782& 0.130422& 1.306130& 0.031\\           
		H(1C) & 0.114630 &0.250000& 1.387644& 0.055\\	
		H(1D) &0.082072& 0.380335 &1.300870& 0.055\\
		H(1E) &0.082072 &0.119665& 1.300870 &0.055\\		
		H(2A) & 0.272797 &0.130407& 1.027536& 0.035 \\	
		H(2B) & 0.272797& 0.369593& 1.027536& 0.035\\
		H(3A) & 0.176232 &0.380411& 1.161826& 0.040 \\	
		H(3B) & 0.176232 &0.119589& 1.161826& 0.040\\
		H(4A) & 0.331681& 0.380527& 1.171267& 0.039\\	
		H(4B) & 0.331681& 0.119473& 1.171267& 0.039 \\
		H(6A) &0.389425& 0.250000 &0.944118& 0.065\\	
		H(6B) &0.423344 &0.380335& 1.030336& 0.065\\
		H(6C) &0.423344& 0.119665& 1.030336& 0.065\\
		\hline\hline
	\end{tabular}
\end{table*}

\begin{table*}[h!]
	\setlength{\tabcolsep}{0.2cm}
	\caption{The atomic coordinates ($x,y,z$) for (C$_4$H$_{14}$N$_2$)Cu$_2$Br$_6$. $U_{\rm iso}$ is the isotropic atomic displacement parameters which is defined as one-third of the trace of the orthogonal $U_{\rm ij}$ tensor. The errors are from the least-square structure refinement. The positions of hydrogen atoms are fixed.}
	\label{atmic position_Br}
\begin{tabular}{ccccccc}
			\hline \hline
			Atomic sites & $x$ & $y$ & $z$ & $U_{\rm iso}$(\AA$^{2}$) & \\\hline
			   Cu(1) & 0.42529(17) & 0.250000 & 0.17793(18) & 0.0305(9)\\
			   Cu(2) & 0.58099(17) & 0.250000 &-0.00147(17) & 0.0264(8)\\
		       C(1) & 0.7023(16) & 0.250000 & 0.6479(15) & 0.061(6)\\            
			   C(2) & 0.7909(16) &0.250000 & 0.6813(14) & 0.061(6)\\            
			   C(3) &0.6098(15) & 0.250000 & 0.5153(16) & 0.061(6)\\          
			   C(4) & 0.8851(16) & 0.250000 & 0.8141(17)& 0.061(6)\\                   
			   N(1) & 0.6973(11) &0.250000 & 0.5495(11) & 0.032(5)\\    
			   N(2) &0.7953(12) & 0.250000 & 0.7780(12) & 0.031(5)\\                
			   Br(1) & 0.27188(15)& 0.250000& 0.18084(17)& 0.0304(7) \\
			   Br(2) & 0.44217(17)& 0.250000& 0.33685(16)& 0.0402(8) \\      
			   Br(3) & 0.57656(13)& 0.250000& 0.16124(15)& 0.0255(7)\\
			   Br(4) &0.42571(12)& 0.250000& 0.01168(14)& 0.0206(7)\\           
			   Br(5) & 0.57318(14)& 0.250000& -0.15956(15)& 0.0279(7) \\
			   Br(6) & 0.73371(15)& 0.250000& 0.00446(15)& 0.0287(7)\\          
			   H(1A) & 0.723934 & 0.363809& 0.528683& 0.039\\  
			   H(1B ) & 0.723934& 0.136191& 0.528683& 0.039\\           
			   H(1C) & 0.673414& 0.125898& 0.670672& 0.074\\	
			   H(1D) & 0.673414& 0.374102& 0.670672& 0.074\\	
			   H(2A) & 0.819889 &0.125860 &0.658667 &0.074\\	
			   H(2B) & 0.819889& 0.374140& 0.658667& 0.074\\
			   H(2C) & 0.768478 &0.136205 &0.798607 &0.038\\	
			   H(2D) & 0.768478& 0.363795& 0.798607& 0.038\\
			   H(3A) &0.610625& 0.250000 &0.451238& 0.092\\	
			   H(3B) & 0.580928& 0.125891& 0.536182& 0.092\\
			   H(3C) & 0.580928& 0.374109& 0.536182 &0.092\\
			\hline\hline
		\end{tabular}
\end{table*}

\begin{table*}[h!]
	\setlength{\tabcolsep}{1.2cm}
	\caption{Some selected bond lengths for (C$_4$H$_{14}$N$_2$)Cu$_2X_6$ ($X$ = Cl, Br).}
	\label{bond length}
	\begin{tabular}{ccccccc}
		\hline \hline
		(C$_4$H$_{14}$N$_2$)Cu$_2$Cl$_6$&Bond length &(C$_4$H$_{14}$N$_2$)Cu$_2$Br$_6$&Bond length\\
		&(\AA) & & (\AA) \\\hline
		Cu(1)-Cl(1)&2.2724(10)&Cu(1)-Br(1)&2.432(3)\\
		Cu(1)-Cl(2)&2.2449(9)&Cu(1)-Br(2)&2.398(3)\\
		Cu(1)-Cl(3)&2.3647(9)&Cu(1)-Br(3)&2.410(3)\\
		Cu(1)-Cl(4)&2.2897(10)&Cu(1)-Br(4)&2.492(3)\\
		Cu(2)-Cl(3)&2.3447(9)&Cu(2)-Br(3)&2.440(3)\\
		Cu(2)-Cl(4)&2.2832(9)&Cu(2)-Br(4)&2.469(3)\\
		Cu(2)-Cl(5)&2.2343(9)&Cu(2)-Br(5)&2.373(3)\\
		Cu(2)-Cl(6)&2.2740(10)&Cu(2)-Br(6)&2.422(3)\\		        N(2)-C(4)&1.475(4)&C(2)-N(2)&1.45(3)\\
	C(3)-C(4)&1.502(4)&C(2)-C(1)&1.491(18)\\
	N(2)-C(6)&1.472(5)&C(4)-N(2)&1.523(18)\\
	N(1)-C(3)&1.471(4)&N(1)-C(3)&1.479(17)\\
	N(1)-C(1)&1.476(5)&N(1)-C(1)&1.48(3)\\	
		\hline\hline 
	\end{tabular}
	\begin{tablenotes}
		\item Symmetry transformations used to generate equivalent atoms of table~\ref{bond length}:
		 $^1$x, y, z $^2$-x+1/2, -y, z+1/2 $^3$x+1/2, -y+1/2, -z+1/2 $^4$-x, y+1/2, -z $^5$-x, -y, -z $^6$ x-1/2, y, -z-1/2 $^7$-x-1/2, y-1/2, z-1/2 $^8$ x, -y-1/2, z 
	\end{tablenotes}
\end{table*} 

\begin{table*}[h!]
	\setlength{\tabcolsep}{1.2cm}
	\caption{Some selected bond angles for (C$_4$H$_{14}$N$_2$)Cu$_2X_6$ ($X$ = Cl, Br).}
	\label{bond angle}
	\begin{tabular}{ccccccc}
		\hline \hline
	(C$_4$H$_{14}$N$_2$)Cu$_2$Cl$_6$ & Bond angle &(C$_4$H$_{14}$N$_2$)Cu$_2$Br$_6$ &Bond angle \\
		&(\degree) & & (\degree) \\\hline
	    C(3)-N(1)-C(1)&112.8(3)&Cu(2)-Br(4)-Cu(1)&94.73(11)\\
		C(6)-N(2)-C(4)&112.6(3)&Cu(1)-Br(3)-Cu(2)&97.60(12)\\ 
	    N(1)-C(3)-C(4)&112.0(3)&Br(5)-Cu(2)-Br(6)&95.09(12)\\ 
	    N(2)-C(4)-C(3)&110.9(3)&Br(5)-Cu(2)-Br(3)&175.36(15)\\  
		Cl(2)-Cu(1)-Cl(1)&95.97(4)&Br(6)-Cu(2)-Br(3)&89.54(11)\\  
		Cl(2)-Cu(1)-Cl(4)& 90.60(4)&Br(5)-Cu(2)-Br(4)&91.59(11)\\ 
	Cl(1)-Cu(1)-Cl(4)&173.43(4)&Cl(2)-Cu(2)-Br(4)&173.32(13)\\ 
	Cl(2)-Cu(1)-Cl(3)&173.37(4)&Br(3)-Cu(2)-Br(4)&83.77(10) \\
	Cl(1)-Cu(1)-Cl(3)&90.66(3)&	Br(2)-Cu(1)-Br(3)&89.55(12) \\
	Cl(4)-Cu(1)-Cl(3)&82.77(3)&	Br(2)-Cu(1)-Br(1)&95.37(13) \\
	Cl(5)-Cu(2)-Cl(6)&94.43(4)&	Br(3)-Cu(1)-Br(1)&175.07(15)\\ 
	Cl(5)-Cu(2)-Cl(4)&175.30(4)&Br(2)-Cu(1)-Br(4)&173.44(16)\\ 
	Cl(6)-Cu(2)-Cl(4)&90.26(3)&Br(3)-Cu(1)-Br(4)&83.89(11)\\ 
	Cl(5)-Cu(2)-Cl(3)&91.95(3)&	Br(1)-Cu(1)-Br(4)&91.18(12)\\ 	
	Cl(6)-Cu(2)-Cl(3)&173.62(4)&N(2)-C(2)-C(1)&112(2)\\ 
	Cl(4)-Cu(2)-Cl(3)&83.36(3)&	C(2)-N(2)-C(4)&114(2)\\ 
	Cu(2)-Cl(3)-Cu(1)&95.03(3)&	C(3)-N(1)-C(1)&113.4(16)\\ 
	Cu(2)-Cl(4)-Cu(1)&98.84(4)&	N(1)-C(1)-C(2)&113(2)\\ 
    \hline\hline 
	\end{tabular}
	\begin{tablenotes}
		\item Symmetry transformations used to generate equivalent atoms of table~\ref{bond angle}:
		 $^1$x, y, z $^2$-x+1/2, -y, z+1/2 $^3$x+1/2, -y+1/2, -z+1/2 $^4$-x, y+1/2, -z $^5$-x, -y, -z $^6$ x-1/2, y, -z-1/2 $^7$-x-1/2, y-1/2, z-1/2 $^8$ x, -y-1/2, z 
	\end{tablenotes}
\end{table*} 

\begin{table*}
	\setlength{\tabcolsep}{.6cm}
	\caption{ Anisotropic atomic displacement parameters (\AA$^{2}$) of (C$_4$H$_{14}$N$_2$)Cu$_2$Cl$_6$.}
	\label{Uaniso_Cl}
	\begin{tabular}{ccccccc}
		\hline \hline
		Atomic sites&$U^{11}$ &$U^{22}$& $U^{33}$ &$U^{23}$ & $U^{13}$&$U^{12}$ \\\hline
		Cu(1)&0.0196(2)&0.0393(3)&0.0188(2)&0.000&0.00149(17)&0.000\\
		Cu(2)&0.0189(2)&0.0310(3)&0.0191(2)&0.000&0.00134(17)&0.000\\
		C(1)&0.031(2)&0.049(3)&0.029(2)&0.000&0.0023(17)&0.000\\
		C(3)&0.035(2)&0.038(2)&0.0266(19)& 0.000&-0.0062(17)&0.000\\
		C(4)&0.030(2)&0.043(2)&0.0240(18)& 0.000&-0.0048(16)&0.000\\
		C(6)&0.034(2)&0.056(3)&0.041(2)&0.000&0.0053(19)&0.000\\
		N(1)&0.0259(17)&0.0266(17)&0.0244(15)&0.000&-0.0054(13)&0.000\\
		N(2)&0.0315(18)& 0.0304(18)& 0.0248(16)& 0.000 &-0.0032(13)& 0.000\\
		Cl(1) & 0.0199(5)& 0.0272(5)& 0.0378(5)& 0.000& 0.0048(4)& 0.000\\
		Cl(2) & 0.0405(6)& 0.0466(6)& 0.0204(4)& 0.000 &0.0002(4)& 0.000 \\      
		Cl(3) &0.0190(5)& 0.0304(5)& 0.0191(4)& 0.000& -0.0004(3)& 0.000\\
		Cl(4) &0.0213(5)& 0.0376(5)& 0.0213(4)& 0.000& -0.0014(3)& 0.000\\           
		Cl(5) &0.0287(5)& 0.0369(5)& 0.0186(4)& 0.000& 0.0026(3)& 0.000\\
		Cl(6) & 0.0188(5)& 0.0302(5)& 0.0338(5)& 0.000& 0.0008(4)& 0.000\\          
			\hline\hline
	\end{tabular}
\end{table*}

\begin{table*}
	\setlength{\tabcolsep}{.6cm}
	\caption{ Anisotropic atomic displacement parameters (\AA$^{2}$) of (C$_4$H$_{14}$N$_2$)Cu$_2$Br$_6$}
	\label{Uaniso_Br}
	\begin{tabular}{ccccccc}
		\hline \hline
		Atomic sites&$U^{11}$ &$U^{22}$& $U^{33}$ &$U^{23}$ & $U^{13}$&$U^{12}$ \\\hline
		Cu(1) & 0.0213(15)& 0.051(2)& 0.0190(15)& 0.000& 0.0015(10)& 0.000\\
		Cu(2) & 0.0187(15)& 0.0407(19)& 0.0197(15)& 0.000& 0.0006(9)& 0.000\\
		C(1) & 0.132(18)& 0.023(8)& 0.029(7)& 0.000& 0.029(10) &0.000\\            
		C(2) & 0.132(18)& 0.023(8)& 0.029(7)& 0.000& 0.029(10) &0.000\\              
		C(3) & 0.132(18)& 0.023(8)& 0.029(7)& 0.000& 0.029(10) &0.000\\           
		C(4) & 0.132(18)& 0.023(8)& 0.029(7)& 0.000& 0.029(10) &0.000\\                   
		N(1) & 0.032(11)& 0.047(13)& 0.018(9)& 0.000& 0.008(8)& 0.000\\    
		N(2) &0.037(11)& 0.033(12)& 0.025(10)& 0.000& 0.007(9)& 0.000\\                
		Br(1) & 0.0215(12) &0.0290(14) &0.0406(14) &0.000 &0.0105(9)& 0.000 \\
		Br(2) & 0.0463(16)& 0.0529(18)& 0.0213(13)& 0.000& 0.0043(10) &0.000\\      
		Br(3) & 0.0180(12) &0.0337(15) &0.0247(12) &0.000 &-0.0052(8)& 0.000\\
		Br(4) &0.0120(12)& 0.0289(14)& 0.0211(11)& 0.000 &0.0012(7)& 0.000\\           
		Br(5) & 0.0264(12) &0.0383(15) &0.0190(11) &0.000 &0.0020(8)& 0.000 \\
		Br(6) & 0.0170(12)& 0.0341(15)& 0.0350(14)& 0.000 &-0.0001(8)& 0.000\\          
		\hline\hline
	\end{tabular}
\end{table*}

\section{Bond Length Distortion and Bond Angle Variance Calculations}
The additional structural parameters such as atomic coordinates, bond lengths, bond angles, and anisotropic atomic displacement parameters are tabulated in Tables~\ref{atmic position_Cl}, \ref{atmic position_Br}, \ref{bond length}, \ref{bond angle}, \ref{Uaniso_Cl}, and \ref{Uaniso_Br}.
The Cu-$X$ bond distances in the basal plane of the Cu$X_6$ units varies between 2.234 to 2.365~\AA~for CuCl$_6$ whereas the apical Cu-Cl bond distances are in the range of 3.014 to 3.022~\AA. Similarly for the CuBr$_6$ units, the Cu-Br bonds of the basal plane lie between 2.373 to 2.492~\AA~ and apical Cu-Br bond distances are in the range of 3.163 to 3.169~\AA. The $\angle X$-Cu-$X$ bond angles are in the range between 82.77 to 95.97$^{\degree}$ in case of CuCl$_6$ units and that for CuBr$_6$ units vary between 83.77 to 95.38$^{\degree}$ in the basal plane.

\begin{figure*}
	\includegraphics[scale=0.9]{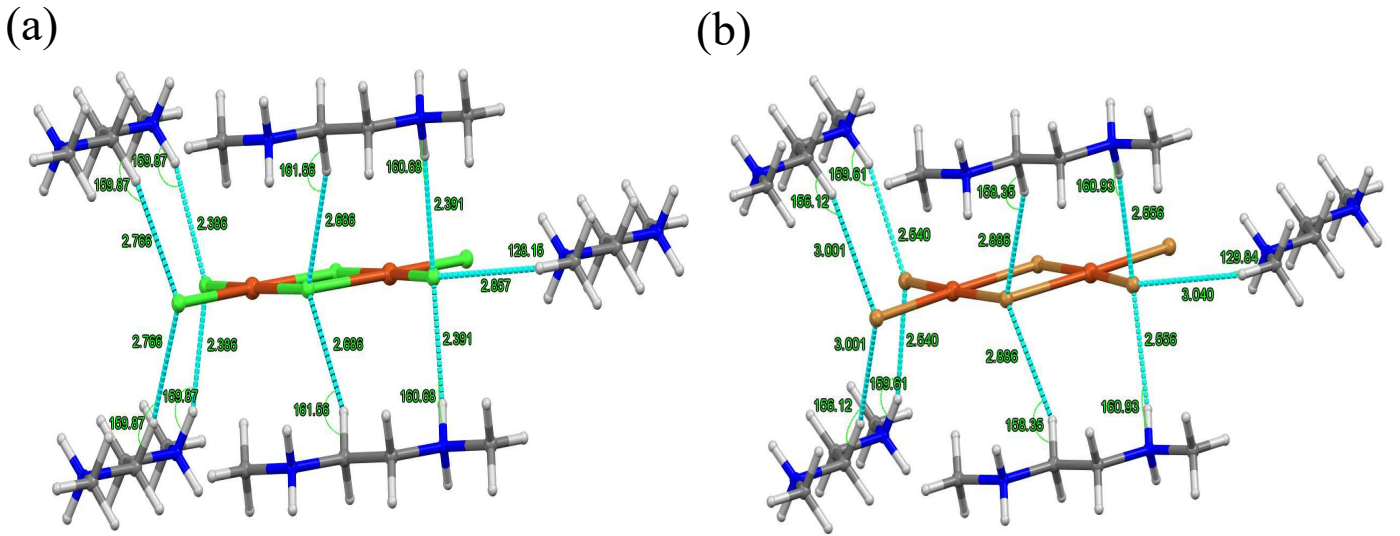}
	\caption{(a) and (b) represent N-H···$X$ and C-H···$X$ hydrogen bonds between organic cation and inorganic anions, $X$ = Cl and Br for (C$_4$H$_{14}$N$_2$)Cu$_2$Cl$_6$ and (C$_4$H$_{14}$N$_2$)Cu$_2$Br$_6$, respectively.}
	\label{Fig1_Supp}
\end{figure*}

As the bond lengths and bond angles are significantly away from the regular octahedral geometry, we have used bond length distortion $\Delta d=\left(\frac16\right)\sum_n\left[\frac{d_n-d_{\rm av}}{d_{\rm av}}\right]^2$ and bond angle variance $\sigma_{\rm oct}^2=\left(\frac1{11}\right)\sum_{i=1}^{12}[\alpha_i-90]^2$ equations to quantify the distortions [$d_{\rm n}$ and $d_{\rm av}$ are the individual and average Cu-$X$ bond lengths and $\alpha_i$ are the individual $\angle X$-Cu-$X$ bond angles]~\cite{Lu13030}. $\Delta d$ and $\sigma_{\rm oct}^2$ values for Cu$_2$Cl$_{10}$ are found to be $1.85\times10^{-2}$ [for Cu(1)], $1.87\times10^{-2}$ [for Cu(2)] and 20.01 [for Cu(1)], 11.57 [for Cu(2)], respectively, whereas the same quantities are found to be $1.69\times10^{-2}$ [for Cu (1)], $1.7\times10^{-2}$ [for Cu(2)] and 15.27 [for Cu(1)], 10.24 [for Cu(2)] for Cu$_2$Br$_{10}$, indicating CuCl$_6$ octahedra are more distorted compared to CuBr$_6$ octahedra. In Cu$_2X_{10}$ dimers, the Cu-Cu distance is 3.473~\AA~for Cu$_2$Cl$_{10}$ and that for Cu$_2$Br$_{10}$ is about 3.650~\AA~in the basal plane, a slight increase in the Cu-Cu distance for later can be attributed to the larger ionic radius for the bromide ion compared to the chloride ion. The nearest Cu$_2X_{10}$ dimers are arranged in an orthogonal manner in the $ac$-plane of the crystal. The interstitial space are filled by the organic cations and interact with the inorganic Cu$_2X_{10}$ dimer via hydrogen bond. A network of strong N-H···Cl and N-H···Br hydrogen bonds between the organic cation and the inorganic anion stabilize both the structures as shown in Fig.~\ref{Fig1_Supp}.

\section{Thermogravimetric Analysis (TGA)}
\begin{figure*}
	\includegraphics[scale=1.7]{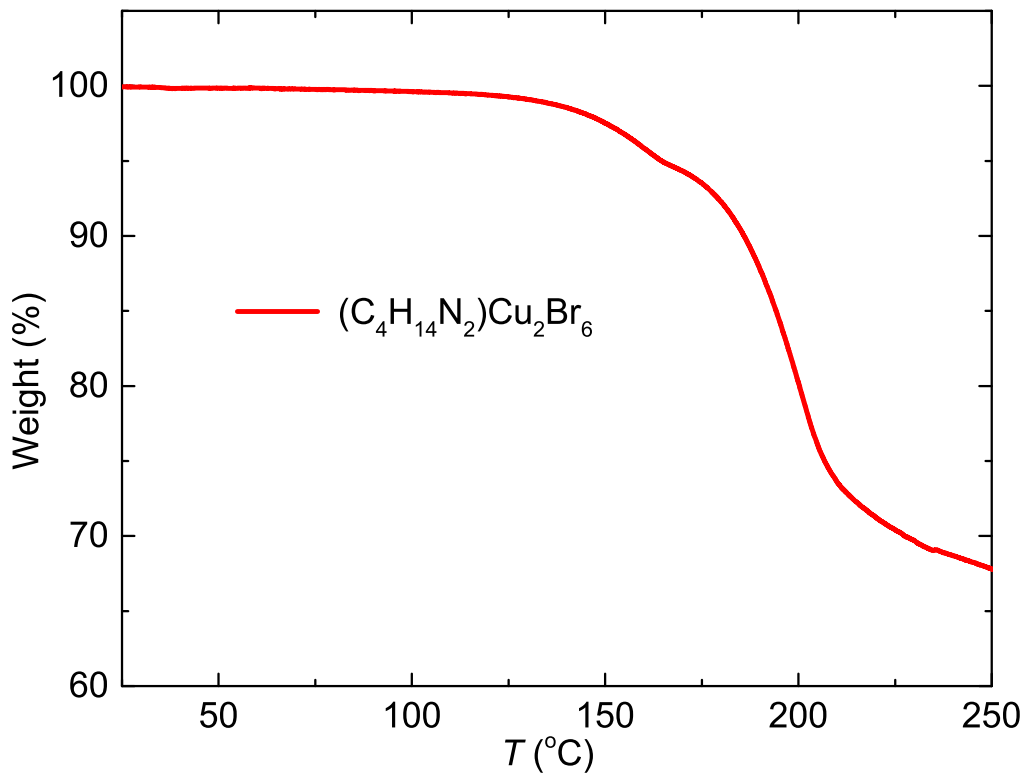}
	\caption{Thermogravimetric analysis of (C$_4$H$_{14}$N$_2$)Cu$_2$Br$_6$.}
	\label{Fig2_Supp}
\end{figure*}

The thermal stability of (C$_4$H$_{14}$N$_2$)Cu$_2$Br$_6$ was checked by thermogravimetric analysis (TGA) at a heating rate of 5 K/min. Unfortunately, this TGA shows that this compound is stable only up to around 420~K (see Fig.~\ref{Fig2_Supp}). In the TGA data, there was 30\% of weight loss between 420 and 500~K.


\end{document}